\documentclass[%
 aip,
 amsmath,amssymb,
  reprint,%
  floatfix
]{revtex4-1}

\usepackage{graphicx}
\usepackage{dcolumn}
\usepackage{bm}

\usepackage[utf8]{inputenc}
\usepackage[T1]{fontenc}
\usepackage{mathptmx}

\begin{document}

	\title{Nonlocal diffusion currents at the nanoscale} 

\author{V. I. Tokar}
\altaffiliation[Also at ]{G. V. Kurdyumov Institute for Metal Physics of the N.A.S. of Ukraine.}
\affiliation{Universit\'e de Strasbourg, CNRS, IPCMS, UMR 7504,
F-67000 Strasbourg, France}
\date{\today}

\begin{abstract}
An integro-differential expression for the diffusion current of the
impurities diffusing by the mechanism of bound impurity-defect pairs
has been derived. The ensuing nonlocal diffusion equation generalizes
the existing theories of diffusion by the pair mechanism in unbounded
systems with homogeneous defect distributions on the systems with
boundaries and variable defect concentration. It has been established
that the nonlocality manifests itself only at short diffusion times
while at large times, in particular, under the stationary conditions,
the predictions of the nonlocal theory would coincide with existing
approaches based on local diffusion currents.  Possibilities of
experimental verification of the nonlocal theory have been suggested.
In particular, the explanation of the uphill diffusion in silicon
by the pair drag can be tested within slightly modified conventional
experimental setups. Also, it has been shown that in the nonlocal theory
the impurity segregation profile caused by the drag should differ at the
initial stage of evolution from the predictions of the local theories.
Because the segregation influences mechanical, corrosive, and electronic
properties of materials, the nonlocal character of the pair diffusion
may have important implications for nanostructured materials.
\end{abstract}

\maketitle
\section{Introduction}
In crystalline solids the substitutional solute atoms diffuse
predominantly by indirect mechanisms with the mobility of the impurities
(the solute atoms in dilute alloys) induced by the mobile point
defects,---the vacancies (v) and the interstitials (I).\cite{philibert,rmp}
Under such mechanisms the diffusion can be strongly influenced by
the impurity-defect (i-d) interaction.  For example, in the case
of strong i-d attraction the tightly bound complexes (or pairs)
form\cite{lidiard,krivoglaz1961} which may considerably enhance
the impurity mobility because the diffusion-mediating defect is
always available within the pair. Besides, the pair diffusion
may underlie the solute segregation at the grain boundaries in
quenched\cite{grain_bound1968,uphill1968-2} and/or irradiated
metals.\cite{irradiation1976,Wiedersich83,radiation2016} The excess
defects flowing to the boundary sinks may drag along the bound impurities
and if the defect gradients are sufficiently large the impurities will
diffuse uphill, that is, from small to large solute concentrations,
thus causing their pileup near the boundary.

Segregation at the grain boundaries and the external surfaces is of great
practical interest because it may significantly impact the mechanical and
corrosive properties of metals (see Ref.\ \onlinecite{radiation2016} and
references therein) and also the electronic properties of semiconductor
devices.\cite{rmp,cowern2007} In modeling the segregation, various
theoretical techniques have been used such as the phenomenological rate
equations,\cite{irradiation1976,Wiedersich83,radiation2016,rmp,cowern1990},
generalized Fick's laws and the Onsager
formalism,\cite{drag-inBCC,radiation2016} as well as the Monte
Carlo\cite{Fe-P-v,SOISSON2006235} and the molecular dynamic
simulations.\cite{P-Fe2005,P-Fe2019} 

The theoretical studies revealed that in the systems with high impurity
concentration and highly supersaturated defects of both kinds the
pair diffusion represents only one of many diffusion routes which
hinders its detailed investigation in such systems.  To facilitate
the task, it seems reasonable to resort to model systems where (i)
the pair mechanism dominates the diffusion, (ii) is driven by a
small concentration of defects of only one kind: the vacancy or the
interstitial, and (iii) in dilute alloys where the i-i interactions can
be neglected.  This idealized picture was approximately realized in
Refs.\ \onlinecite{cowern1990,cowern1991,cowern1992} in experiments
on diffusion of boron impurities in silicon. The simplicity of
experimental setup allowed the authors to describe the diffusion with
only two rate equations which made possible a detailed analysis of the
pair mechanism. The most important outcome of the analysis of Ref.\
\onlinecite{cowern1990} was that the pair diffusion is intrinsically
non-Fickian (n-F). The reason is that when the pair nucleates at
some point in the crystal and then diffuses on a macroscopic distance
away from it, the impurity current carried by the pair is defined to
a large extent by the densities of the impurities and the defects at
the initial point but is not related to the local concentrations at the
actual position of the pair in contradiction to the first Fick's law.
The nonlocality of the pair currents was not noted in the segregation
studies\cite{irradiation1976,SOISSON2006235,drag-inBCC,radiation2016}
though it may have important experimental consequences, for example,
produce n-F exponential diffusion profiles.\cite{cowern1990}

The aim of the present paper is to derive a phenomenological expression
for the nonlocal current arising in the diffusion by the pair mechanism
and to use it to extend the description of the pair diffusion developed in
Refs.\ \onlinecite{cowern1990,In/cu(001)} for systems with constant defect
concentration to the case of bounded systems with inhomogeneous defect
distribution. This extension will make possible the description of the
uphill diffusion in silicon\cite{uphill0,cowern2007} by the same pair drag
mechanism as in metals.\cite{grain_bound1968,uphill1968-2} Furthermore,
it will predict the possibility of observing the exponential diffusion
profiles seen in the homogeneous systems\cite{cowern1990,In/cu(001)}
in the impurity segregation profiles caused by the drag in strongly
varying defect distributions.
\section{Diffusion by the mechanism of mobile i-d pairs}
To begin with let us qualitatively discuss the description of
the pair diffusion based on the local currents. In the continuum
approximation the pair density at point ${\bf r}$ at time $t$ should be
proportional to the local concentrations of the impurities and of the
defects\cite{irradiation1976,local-current1,local-current2}
\begin{equation}
	C_p({\bf r},t)\propto C_i({\bf r},t)C_d({\bf r},t)
	\label{Cp-CiCd}
\end{equation}
where the proportionality constant is given by i-d pairing
rate.\cite{local-current1,local-current2} In the continuum
approximation the diffusivity of a tightly bound pair can
be characterized by the diffusion constant $D_p$ which in
microscopic models of the vacancy-mediated diffusion can be calculated
explicitly.\cite{lidiard,irradiation1976,krivoglaz-repetskiy,drag-inBCC,I}
Because we assume that the impurity diffuses only within the
pairs, the impurity current should be proportional to the pair
current\cite{local-current2}
\begin{equation}
	{\bf J}_i^{(loc)}\propto-D_p\nabla C_p
	=-\frac{D_i}{C_d^*}C_iC_d(\nabla\ln C_i+\nabla C_d)
	\label{J_local0}
\end{equation}
where $D_i$ and $C_d^*$ are the equilibrium impurity diffusion constant
and the defect concentration, respectively. As is seen, when $C_d({\bf
r},t)=C_d^*$ the standard first Fick's law is recovered. But when $C_d$
depends on the space coordinates and its logarithmic gradient is opposite
to the logarithmic gradient of $C_i$ and, besides, exceeds it in the
absolute value---the current in Eq.\ (\ref{J_local0}) flows along the
impurity gradient causing the uphill diffusion.

In the dilute limit Eq.\ (\ref{J_local0}) agrees with the strong
binding limit of the expression derived in the local approach in Ref.\
\onlinecite{drag-inBCC} for the vacancy-mediated diffusion ($d=v$). 
The generalized Fick's law was expressed in terms of the Onsager matrix
$\hat{L}$ as\cite{drag-inBCC} 
\begin{equation}
	{\bf J}_i^{(loc)}=-L_{ii}\frac{\nabla \mu_{i}}{k_BT}
	-L_{iv}\frac{\nabla \mu_{v}}{k_BT}.
	\label{Onsager}
\end{equation}
Taking into account that in the dilute limit $\nabla \ln C_{i(v)}
= {\nabla \mu_{i(v)}}/{k_BT}$\cite{NpNf} and substituting
this into (\ref{J_local0}) one obtains Eq.\ (\ref{Onsager}) with
$L_{ii}=L_{iv}=D_iC_iC_v/C_v^*$ so that $L_{iv}/L_{ii}=1$ in accordance
with the strong-binding limit in Ref.\ \onlinecite{drag-inBCC}. It is
to be noted that besides Refs.\ \onlinecite{local-current2,drag-inBCC}
the local expression Eq.\ (\ref{J_local0}) can be obtained from
the pair contribution to the diffusion current derived in Refs.\
\onlinecite{irradiation1976,Wiedersich83,SOISSON2006235}.
\subsection{Nonlocal currents}
A distinct consequence of the n-F nature of diffusion by the
pair mechanism is that the impurity profiles observed in Ref.\
\onlinecite{cowern1990} and later in Ref.\ \onlinecite{In/cu(001)}
were not Gaussian as in the Fickian diffusion but had the exponential
shapes. In the latter reference it was shown that this was a direct
consequence of the instability of the pairs which can be shown as
follows. The kernel $G_p$ describing the diffusion of an unstable pair
decaying with the rate $\epsilon$ satisfies the equation\cite{In/cu(001)}
\begin{equation}
	\frac{\partial}{\partial t} G_p({\bf r,r}_0,t)= D_p\nabla^2 
G_p({\bf r,r}_0,t) -\epsilon G_p({\bf r,r}_0,t),
	\label{gastel_eq}
\end{equation}
where ${\bf r}_0$ is the point at which the pair was nucleated at time $t=0$ 
so the initial condition for $G_p$ reads
\begin{equation}
G_p({\bf r,r}_0,t)=\delta({\bf r-r}_0).  
\label{Gp-ini}
\end{equation}
In a homogeneous system the solution of Eq.\ (\ref{gastel_eq}) is given
by the conventional Gaussian kernel multiplied by $\exp(-\epsilon
t)$ to account for the decay.  As $t\to\infty$ the pair completely
dissociates and the impurity atom immobilizes in the substitutional
positions according to the probability distribution given by the
integral\cite{In/cu(001)}
\begin{equation}
P({\bf r,r}_0)=\epsilon \int_0^\infty G_p({\bf r,r}_0,t)dt
\label{Pdef}
\end{equation}
which in all physical dimensions and geometries has the exponential
tail\cite{cowern1990,In/cu(001),I}  
\begin{equation}
P({\bf r,r}_0)|_{|{\bf r-r}_0|\to\infty}\propto\exp(-|{\bf r-r}_0|/\lambda),
\label{Ptail}
\end{equation}
where 
\begin{equation}
\lambda=\sqrt{D_p/\epsilon}
\label{lambda}
\end{equation}
is the average migration distance of the impurity during one i-d
encounter.\cite{cowern1990,In/cu(001)}

Because both the pair diffusion and the decay are determined by the
movements of the defect, the kinetics described by (\ref{gastel_eq})
proceed at the microscopic time scale. The movements of the impurity
at this sale can be so fast that sometimes even cannot be detected
experimentally.\cite{In-V-attraction} In practical applications,
however, one is usually interested not in the diffusion of individual
impurities and/or pairs but in the evolution of the concentration profile
$C_i({\bf r},t)$.  Its evolution takes place at a slow macroscopic
time scale because the speed of the profile change is limited by the
concentration of lattice defects $C_d$ which is usually small due to their
high creation energy in the range of several eV.\cite{rmp,philibert}
Therefore, the substitutional impurity remains immobile during long
intervals between the i-d encounters which trigger short bursts of the
diffusion steps.\cite{In-V-attraction} The average time span between the
encounters $t_p=g^{-1}$, where $g\ll1$ is the i-d pairing rate which in
the local approximation is proportional to $C_d$\cite{cowern1991}
\begin{equation}
g({\bf r},t)=g^*C_d({\bf r},t)/C_d^*
\label{g-Cd}
\end{equation}
where the starred quantities are the equilibrium values.  In the
absence of i-d attraction the encounter may amount to only one
impurity displacement on the distance of order of the lattice constant
$a$. However, under strong attraction when the decay rate $\epsilon$
in Eq.\ (\ref{lambda}) is small the bound pair may perform a large
number of the elementary diffusion steps and migrate at a distance
$\lambda\gg a$.\cite{cowern1990,In-V-attraction} In this case the
effective microscopic time scale is defined by the pair lifetime
$\tau=\epsilon^{-1}$ and the effective diffusion step $\lambda$.

If one is interested only in macroscopic evolution at the time scale
$t=O(t_p)\gg\tau$, the infinitesimal time step $\Delta t$ in the governing
equation can be chosen in such a way that it was microscopically large 
but still small at the macroscopic scale:
\begin{equation}
t=O(t_p)\gg\Delta t\gg \tau.
\label{delta_t}
\end{equation}
For example, in the experiments in Refs.\ \onlinecite{In-V-attraction}
the average time interval between the jumps was $10^{-8}$~s and during
the pair lifetime the impurity made on average $\sim10$ steps so $\tau$
was of $O(10^{-7}$~s). The mean time between the pairings, on the other
hand, was $\sim10$~s, so with the choice of $\Delta t=10^{-3}$~s both
inequalities in (\ref{delta_t}) are easily satisfied.  Because of the
four orders of magnitude difference between $\Delta t$ and $\tau$ in this
example, the creations and decays of the pair will be confined within
the time step $\Delta t$ with the relative error $O(10^{-4})$. So on the
macroscopic time scale the redistribution of the impurity density due
to the pair diffusion looks as instantaneous.  This means that within
$\Delta t$ the impurity distribution after one I-d encounter can be
approximated by $P$ in Eq.\ (\ref{Pdef}).\cite{In/cu(001)}

Thus, on the macroscopic scale the impurity diffusion looks like its
disappearance at point ${\bf r}_0$ due to the pairing with a defect
followed by its immediate redistribution in the vicinity of that point
according to the probability density $P$ in (\ref{Pdef}). By analogy
with Fick's second law, the governing equation for the diffusion by this
mechanism can be obtained as the condition of the impurity conservation as
\begin{eqnarray} 
\frac{\partial}{\partial t} C_i({\bf r},t) 
&=&-g({\bf r},t)C_i({\bf r},t)\nonumber\\ 
&+& \int P({\bf r,r}_0) C_i({\bf r}_0,t)g({\bf r}_0,t)d{\bf r}_0. 
\label{the_eq} 
\end{eqnarray} 
Here the first term on the right hand side (r.h.s.) describes the loss of
impurities by the concentration profile due to the pairings and the second
term describes its replenishment by the decays.  

Because the pair diffusion kernel $G_p$ enters Eq.\ (\ref{the_eq}) only
through $P$ from Eq.\ (\ref{Pdef}), the formalism can be simplified by
dealing directly with $P$ without the resort to $G_p$. This is achieved
by integrating Eq.\ (\ref{gastel_eq}) over $t$ from zero to infinity
with the use of the initial condition Eq.\ (\ref{Gp-ini}) which amounts
to the differential equation for $P$
\begin{equation}
	-\lambda^2\nabla^2_{\bf r}P({\bf r,r}_0)+P({\bf r,r}_0)
=\delta({\bf r-r}_0).
\label{P_eq}
\end{equation} 
It should be supplemented by the boundary conditions which can be derived
from the definition of $P$ in terms of $G_p$ Eq.\ (\ref{Pdef}).  Because
the lattice defect, by definition, cannot exist beyond the lattice, the
pair cannot cross the system boundary $B$ which leads to the zero Neumann
boundary condition for the pair flux through the boundary for $G_p$. By
Eqs.\ (\ref{Pdef}) and (\ref{J_local0}) this translates to $P$ as
\begin{equation}
	{\bf n\cdot\nabla} P({\bf r,r}_0)|_{B}=0,
\label{BC}
\end{equation}
where ${\bf n}$ is the vector normal to the boundary.

With the use of Eq.\ (\ref{P_eq}) Eq.\ (\ref{the_eq}) can be cast in the
form of the second Fick's law as follows.  According to Eq.\ (\ref{P_eq}),
$P$ is equal to the sum of the delta-function and of the Laplacian
multiplied by $\Lambda^2$. Substituting it in Eq.\ (\ref{the_eq}) one gets
\begin{equation}
\frac{\partial}{\partial t} C_i({\bf r},t) = 	
\lambda^2\nabla^2_{\bf r}\int P({\bf r,r}_0)
C_i({\bf r}_0,t)g({\bf r}_0,t)d{\bf r}_0 
=-\nabla_{\bf r}\cdot{\bf J}_i({\bf r}),
	\label{the_eq2}
\end{equation}
where the impurity current 
\begin{equation}
	{\bf J}_i({\bf r})=-\lambda^2\nabla_{\bf r}\int P({\bf r,r}_0)
	C_i({\bf r}_0,t)g({\bf r}_0,t)d{\bf r}_0.
	\label{J}
\end{equation}
As is seen, in contrast to the local expression (\ref{J_local0}) current
in Eq.\ (\ref{J}) is nonlocal. It depends depends on the weighted
concentrations of impurities and the defects (see Eq.\ (\ref{g-Cd}))
within a region of radius $\sim\lambda$ defined by the exponential
attenuation of $P$ with distance (see Eq.\ (\ref{Ptail})). Physically
the region corresponds to the neighborhood of point ${\bf r}$ from which
the impurity may reach the point in one migration event.

When $\lambda$ tends to zero, the current in Eq.\ (\ref{J})
becomes local, as can be seen from (\ref{P_eq}):
\begin{equation}
	P({\bf r,r}_0)|_{\lambda\to0}\to\delta({\bf r-r}_0). 
	\label{lim_P}
\end{equation}
The migration distance,however, is not a tunable parameter but has a
concrete finite value specific to the system. But because it has the
dimension of length, it should be compared with the typical length scale
on which the functions entering the integrand in Eq.\ (\ref{J}) vary.
Assuming that on the scale of $\lambda$ the variation of both $C_i$ and
$g$ is weak, all functions of ${\bf r}$, $P$ in the integrand of Eqs.\
(\ref{the_eq}) and (\ref{the_eq2}) can be replaced by the local expression
(\ref{lim_P}) to give
\begin{equation}
	{\bf J}_i^{(loc)}({\bf r},t)\simeq-\lambda^2\nabla
[C_i({\bf r},t)g({\bf r},t)]
	\label{J_local}
\end{equation}
which in view of Eq.\ (\ref{g-Cd}) coincides with the local current
Eq.\ (\ref{J_local0}). Though in explicit calculations below Eq.\
(\ref{J_local0}) will always be used, notation $g({\bf r},t)$ will be
kept in order to remind that Eq.\ (\ref{g-Cd}) is only an approximation
and in a rigorous treatment $g$ must be calculated with the use of the
diffusion equation for the defects and with an appropriate model for
i-d pairing. In particular, because both the impurity and the defect are
present within the pair, a contribution similar to Eq.\ (\ref{the_eq})
and/or Eq.\ (\ref{the_eq2}) should be accounted for also in the defect
diffusion.  In the present study it is neglected because in contrast
to the impurity the defects are the most mobile entities in the system
so when the impurity concentration is small the pair contribution which
is proportional to $C_i$ is negligible. This approximation, however,
may not be appropriate in concentrated alloys.\cite{local-current2}

Eq.\ (\ref{the_eq2}) together with Eqs.\ (\ref{P_eq}),
(\ref{BC}), and (\ref{the_eq}) are the main results of the present
study. Besides the heuristic derivation presented above, they
agree with the rigorous theory of Ref.\ \onlinecite{drag-inBCC}
in the limit of slow variation of the defect and the impurity
concentrations as well as with the heuristic approaches of Refs.\
\onlinecite{irradiation1976,local-current2,SOISSON2006235}.
The equations extend the picture of the pair diffusion developed in
Refs.\ \onlinecite{cowern1990,In/cu(001)} for unbounded systems with
constant defect concentration to the inhomogeneous case and systems with
boundaries. In particular, the nonlocal equations exactly reproduce
the evolution of 1D impurity concentration profile described in Ref.\
\onlinecite{cowern1990}.  In an unbounded host with constant $g$ Eqs.\
(\ref{P_eq}) and (\ref{the_eq2}) or, equivalently, Eq.\ (\ref{the_eq}),
can be solved with the use of the Fourier transform as follows. The
solution of Eq.\ (\ref{P_eq}) reads
\begin{equation}
	P({\bf k}) = (1+\lambda^2 {\bf k}^2)^{-1},
	\label{Pk}
\end{equation}
where ${\bf k}$ is the Fourier momentum. With known $P({\bf k})$, the 
solution of Eq.\ (\ref{the_eq2}) is easily found to be given by
\begin{equation}
	C_i({\bf k},t)=\exp\left(-\frac{g\lambda^2 {\bf k}^2}
{1+\lambda^2 {\bf k}^2}t\right) C_i({\bf k},t=0).
	\label{Ckt}
\end{equation}
In Appendix \ref{appA} it has been shown that in 1D geometry with the
initial delta-function profile $C_i({\bf k},t=0)=1$, where $k=k_x$
is the $x$-component of ${\bf k}$, the inverse Fourier transform
of Eq.\ (\ref{Ckt}) coincides with the series solution of Ref.\
\onlinecite{cowern1990}.  It is important to stress that this solution,
hence, the underlying physical picture, were experimentally confirmed
in Refs.\ \onlinecite{cowern1990,cowern1991,cowern1992,mirabella} so
the nonlocal equations generalize this picture to a broader class of
the pair diffusion problems.
\section{Impurity diffusion in stationary defect profiles} 
To estimate the uphill diffusion and/or the segregation predicted by the
nonlocal equations let us consider the evolution of an impurity profile
under a stationary inhomogeneous defect distribution. The inhomogeneity
naturally arises, for example, when the concentration of defects in
the crystal bulk exceeds their value at the surface, so in the absence
of other sinks they flow towards it to reduce the supersaturation.
In the case of sufficiently strong i-d attraction the impurity dragged
by the defects will diffuse in the same direction, as follows from Eq.\
(\ref{J_local0}).\cite{grain_bound1968,uphill1968-2,drag-inBCC} As
has been pointed out previously, the drag may take place even when the
impurity concentration in the bulk is lower than near the surface. It
is this uphill diffusion that underlies the alloy segregation by the
pair drag mechanism.\cite{grain_bound1968,uphill1968-2}

The strict stationarity of the defect distribution can
be achieved only in the presence of time-independent
defect sources, such as the stationary external
irradiation\cite{irradiation1976,Wiedersich83,SOISSON2006235,mirabella2009,%
radiation2016} or in an externally imposed constant temperature
gradient.\cite{soret1964} But as will be seen below, of major
interest to the present study is the initial stage of segregation,
so the stationarity should be a reasonable approximation
even in cases of transient defect supersaturation as produced,
e.g., by the quench,\cite{grain_bound1968,uphill1968-2} by the ion
implantation,\cite{rmp,cowern2007} by the oxidation,\cite{cowern1990} etc.
The qualitative behaviors discussed below should also hold in the case
of time-dependent supersaturation because the underlying mechanism of
the pair diffusion will be operative also in this case.

The time-independence of defect distributions significantly facilitate
analysis because in this case the pairing rate $g({\bf r})$ entering
the diffusion equations will also be stationary.  Furthermore, because
diffusion equations describe relaxation processes, at large times the
impurity profiles should also asymptotically approach stationary shapes.
With the use of Eq.\ (\ref{the_eq2}) it can be shown that as $t\to\infty$
the impurity distribution will acquire the form
\begin{equation} 
C_i({\bf r},t\to\infty) \simeq \frac{A}{g({\bf r})}
\label{Cstat} 
\end{equation} 
where $A$ is a normalization constant defined by the total number
of impurities in the system.  The validity of Eq.\ (\ref{Cstat})
can be seen by observing that $P({\bf r,r}_0)$ in Eq.\ (\ref{P_eq})
is a symmetric function of its arguments which is a property of Green's
functions satisfying the Neumann boundary condition Eq.\ (\ref{BC}). So
the integral in Eq.\ (\ref{the_eq2}) over ${\bf r}_0$ is equal to
the integral in Eq.\ (\ref{P_eq}) over ${\bf r}$, i.e., is equal to
unity, which makes the r.h.s.\ of Eq.\ (\ref{the_eq2}) equal to zero.
The peculiarity of the profile Eq.\ (\ref{Cstat}) is that it does not
depend on $\lambda$ which means that it will be the same for both large
and small $\lambda\to0$, that is, in the local theory. In the latter
case substituting Eq.\ (\ref{J_local0}) into the second Fick's law Eq.\
(\ref{the_eq2}) one arrives at the local diffusion equation
\begin{equation}
	\frac{\partial C_i({\bf r},t)}{\partial t}
= \lambda^2\nabla^2[g({\bf r},t)C_i({\bf r},t)]\simeq
	\frac{D_i}{C_d^*}\nabla^2[C_d({\bf r},t)C_i({\bf r},t)]
	\label{diff-eq}
\end{equation}
where the second equality follows from Eq.\ (\ref{g-Cd}) and from the
expression\cite{cowern1990}
\begin{equation}
D_i=g^*\lambda^2.
	\label{D-g}
\end{equation}
As is easily seen, the stationary profile Eq.\ (\ref{Cstat}) satisfies also
the local diffusion equation (\ref{diff-eq}).

Equations similar to Eq.\ (\ref{diff-eq}) are frequently used for
the description of the pair-driven drag. In particular, in the
BCC\cite{drag-inBCC} and in the FCC metals in the vacancy-mediated
diffusion (see equation (30) in Ref.\ \onlinecite{irradiation1976}
in the strong binding limit\cite{philibert} $w_3 \to0$).  In the
case of interstitial defects $d=I$ the last of Eqs.\ (11) in Ref.\
\onlinecite{SOISSON2006235} or the corresponding equation in Ref.\
\onlinecite{Wiedersich83} would acquire the form of Eq.\ (\ref{diff-eq})
if the vacancy contribution were neglected. The theories in these
papers were intended to deal mainly with the drag under the stationary
irradiation and the stationary segregation profiles so the local should be
sufficient.  However, in the next section it will be shown that at small
diffusion times the local form may lead to erroneous conclusions about the
shape of the impurity distributions so more accurate nonlocal equations
should be used in the simulations of the short-time pair diffusion.
\section{\label{profiles}Impurity drag towards the surface by the pairs}
To make a semi-quantitative comparison of the segregation predicted by
the nonlocal diffusion equations with experimental observations and the
local theory, the impurity profiles in a symmetric film of thickness
$2L$ has been simulated.\cite{irradiation1976}  Our interest was in
the profiles that strongly vary at distances comparable to $\lambda$
when the local approximation can break down.

The pairing rate was modeled by a simple stationary shape
\begin{equation}
g(x)=\left\{\begin{array}{lc}
g_s[s^{-1}+(1-s^{-1})x/d_s],\quad  &0\leq x<d_s,\\
g_s,\quad  &d_s\leq x\leq L,
\end{array}\right.
\label{g(x)}
\end{equation}
where $d_s$ is the depth at which the pair nucleation rate saturates
to a constant value $s=g_s/g(0)$ which will be called supersaturation
and assumed to be much larger than unity.\cite{cowern1990} For large
$d_s$ the variation of $g(x)$ in Eq.\ (\ref{g(x)}) will be slow so the
choice of Eq.\ (\ref{g(x)}) can be justified by Eq.\ (\ref{g-Cd}) as
follows. If starting from depth $d_s$ there exist some stationary defect
sources, such as the interstitial clusters,\cite{cowern2007} which keep
a constant level of defect supersaturation for $x\ge d_s$ and if at the
surface the defect concentration drops to much smaller equilibrium value
$g(0)=g^*$,\cite{irradiation1976,cowern2007} the conventional 1D diffusion
equation would predict the linear defect profile in the interval $0\leq
x\leq d_s$ and by Eq.\ (\ref{g-Cd}) also the pairing profile.

The uphill diffusion for a system roughly modeling the boron
diffusion in silicon\cite{cowern2007} has been simulated with the
use of Eqs.\ (\ref{the_eq}) and (\ref{g(x)}). The input parameters
in the simulations have been chosen to be similar to those in Refs.\
\onlinecite{cowern1990,supersaturation,cowern2007}.  The parameters
entering Eq.\ (\ref{g(x)}) were assessed departing from the fact that
in Ref.\ \onlinecite{supersaturation} strong gradients of the defect
concentration were observed below the depth $\sim10$~nm so $d_s$
was set at this value and supersaturation $s=10^2$ has been assumed
as in Ref.\ \onlinecite{cowern1990};  $g_s=sg^*$ was calculated from
the expression for the diffusion constant Eq.\ (\ref{D-g}) assuming
$\lambda\simeq5$~nm.\cite{cowern2007} The equilibrium diffusion constant
was taken from Ref.\ \onlinecite{D0Q_B2000}. The film thickness
$L=200$~nm used in the simulations was chosen to be larger than in
experiments of Ref.\ \onlinecite{cowern2007} in order to exclude the
influence of the second surface of the film and to justify the use of
the simple expression Eq.\ (\ref{Pxx0}) for $P$ valid for $L\gg\lambda$.
This should be adequate for the simulation of impurity pileup near the
surface because the experimental data in this region are very similar both
for the film and for the surface of the bulk material.\cite{cowern2007}

\begin{figure}
\begin{center}
\includegraphics[viewport = 30 20 300 205,scale=0.75]{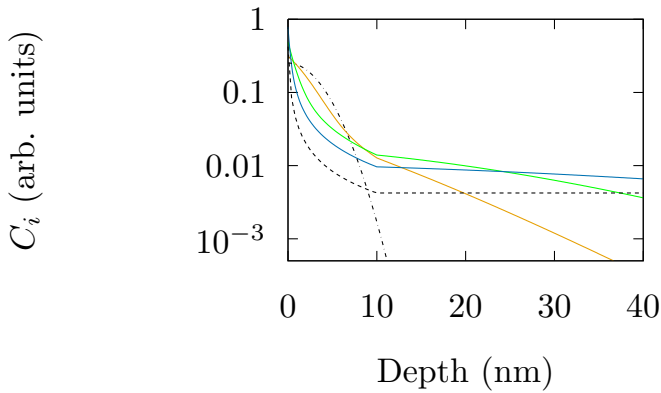}
\end{center}
\caption{\label{fig1}Solid lines: model diffusion profiles with
parameters corresponding to B-Si system simulated with the use of Eqs.\
(\ref{the_eq}), (\ref{g(x)}), and (\ref{Pxx0}).  The diffusion lasted
60~s at $T=$~850, 900, and 950$^\circ$~C. The broader tails and narrower
shapes near the surface correspond to higher temperatures. Dashed line:
the stationary profile Eq.\ (\ref{Cstat}); dashed-dotted line: the
initial Gaussian impurity distribution.}
\end{figure}

The results of simulations shown in Fig.\ \ref{fig1} are similar to the
experimental profiles of Ref.\ \onlinecite{cowern2007} but a slight
difference can be noted.  Namely, the profile corresponding to the
lowest temperature has slightly positive curvature near the surface
which is absent on other two curves and at the limiting curve Eq.\
(\ref{Cstat}). This can be attributed to the fact that with the same
diffusion time (60~s) the smaller temperature means earlier stage
of diffusion so the shape has not yet reached the asymptotic regime.
In experiments of Ref.\ \onlinecite{cowern2007} all three curves are
qualitatively similar which means that they correspond to the advanced
stage of the evolution. Thus, in experimental conditions of Ref.\
\onlinecite{cowern2007} one cannot distinguish between the local
and non-local cases but our order-of-magnitude estimates show that
the conditions are quite close to the preasymptotic regime where this
distinction could be detected.  And it seems that only the nonlocality
and, more importantly, the measurement of $\lambda$ and its comparison
with independently obtained value could definitely prove the pair
mechanism of the uphill diffusion in B-Si system.

However, the simulations also have shown that the experimental setup of
Refs.\ \onlinecite{uphill0,cowern2007} is not very convenient for the
quantitative study of the pair diffusion.  The cause of difficulty lies in
the initial impurity profile which is peaked near the surface (see, e.g.,
Ref.\ \onlinecite{cowern2007}) and due to inevitable experimental errors
in its measurement the additional pileup would be difficult to measure
with sufficient accuracy to distinguish between theoretical predictions
of different models. To reliably measure the diffusion profile it seems
more natural to chose the initially homogeneous impurity distribution
\begin{equation}
C_i({\bf r},0)=C_{i0}=Const
\label{C-const}
\end{equation}
and also to shift $d_s$ further from the surface in order that the
large second derivative of $C_d$ at $d_s$ (see Eq.\ (\ref{diff-eq}))
did not obscure the pair contribution.

\begin{figure}
\begin{center}
\includegraphics[viewport = 30 20 300 205,scale = 0.7]{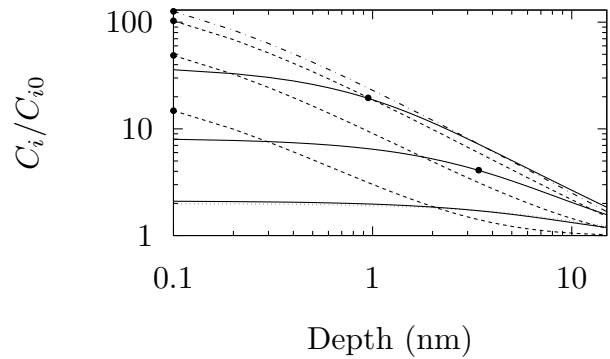}
\end{center}
\caption{\label{fig2}Solid lines: impurity profiles simulated with
the use of Eq.\ (\ref{the_eq}) for (from bottom to top) $\bar{g}t=1$,
5 and 20;  dashed lines: the profiles calculated with the use of Eq.\
(\ref{diff-eq}) for the same parameter sets; The bullets mark the
distributions at half maximum.  Dashed-dotted line is the asymptotic
distribution Eq.\ (\ref{Cstat}) and the dotted line is the approximate
solution Eq.\ (\ref{C_ini}) (for farther explanations see the text.)}
\end{figure}

To this end in the simulations below the boundary between the regions of
the linear and the constant behaviors in Eq.\ (\ref{g(x)}) was shifted
towards $d_s=100$~nm but the gradient of the supersaturation was kept the
same as previously, that is at 10~nm$^{-1}$, which means supersaturation
$s=10^3$ at $d_s$.  Besides, larger value of $\lambda=10$~nm has been
used\cite{cowern1990} to simplify visualization of the profiles in the
region of interest (see Fig.\ \ref{fig2}).

In Fig.\ \ref{fig2} are shown the impurity profiles near the surface
simulated with the use of Eqs.\ (\ref{the_eq}) and (\ref{diff-eq})
with the same diffusion constant $D_i$ and $g(x)$ in Eq.\ (\ref{g(x)})
in both cases. According to Ref.\ \onlinecite{cowern1990}, a natural time
scale for the evolution of the diffusion profile under the pair mechanism
is the number of the migration events of the i-d pairs. When the number
of migrations is equal to unity the diffusion amounts to a single i-d
encounter with the diffusion profile acquiring characteristic exponential
shape. Large number of migrations will lead to the conventional Gaussian
behavior.\cite{cowern1990} However, under the inhomogeneous defect
distribution the pair nucleation rate depends on the distance to the
surface so the probability of the encounter depends on the position of
the impurity within the film so the calculation of the parameter $gt$
introduced in Ref.\ \onlinecite{cowern1990} becomes ambiguous. It seems
reasonable to define instead a characteristic pair nucleation rate
$\bar{g}$ in the near-surface region with the use of Eq.\ (\ref{g(x)})
as the rate at the distance $\sim\lambda$ from the surface
\begin{equation}
	\bar{g}\simeq g_s\frac{\lambda}{d_s}.
	\label{g-bar}
\end{equation}
The diffusion profile for small values of $\bar{g}t\lesssim1$
calculated to this accuracy in Appendix \ref{app} in Eq.\ (\ref{Cxt=0}) is
\begin{equation}
	C_i(x,t)\approx C_{i0}\left(1+\bar{g}te^{-x/\lambda}\right).
	\label{C_ini}
\end{equation} 
This expression approximately agrees with the numerical
solution for $\bar{g}t=1$ (see Fig.\ \ref{fig2}). Noticeable
is the exponential contribution characteristic of the pair
diffusion.\cite{cowern1990,In/cu(001)}

To visualize the details of the diffusion profiles in the narrow
region near the surface, the logarithmic scales have been used in Fig.\
\ref{fig2}.  Because of this, the values at $x=0$ are not shown in the
plots but from the bullets that mark the half width of distributions
it can be seen that the concentrations calculated with the use of Eq.\
(\ref{diff-eq}) are twice as large at $x=0$ as at $x=0.1$~nm while the
profiles simulated with the pair-diffusion mechanism Eq.\ (\ref{the_eq})
at $x=0$ exceed $x=0.1$~nm values only on a few percent. It is to
be noted that the equation has been integrated with the step $\Delta
x=0.1$~nm because anyway the continuous approximation cannot be valid
at distances smaller than the atomic radii of order $O(0.1~\mbox{
nm})$. Thus, though at large diffusion times both mechanisms will produce
the same profile Eq.\ (\ref{Cstat}), at small to intermediate times the
profiles simulated with the local diffusion equation will be much more
peaked near the surface. For example, for the pair diffusion mechanism
the impurity density at the surface is an order of magnitude smaller in
the cases $\bar{g}t=1,5$ and about five time smaller for $\bar{g}t=20$
(at large times the distributions converge).  This may have important
practical consequence for the properties of the near-surface layers.
For example, the onset and the distribution of the impurity precipitates
may significantly differ in the two cases.
\section{Conclusion}
In this paper the problem of impurity diffusion by the mechanism of
mobile i-d pairs\cite{cowern1990} has been considered theoretically
and a phenomenological integro-differential expression for the
impurity diffusion current has been derived on the basis of a simple
physical picture of the pair diffusion. The expression is a nonlocal
generalization of the first Fick's law. Being substituted into the
second law it leads to a nonlocal diffusion equation. Its predictions
differ at the early stages of the diffusion from those of the local
equations that are conventionally used in the studies of the impurity
segregation.\cite{irradiation1976,Wiedersich83,SOISSON2006235,drag-inBCC}

The difference between predictions of the two approaches
has been illustrated by simulations of the uphill diffusion
and the ensuing impurity drag toward the surface under
the conditions of defect supersaturation in the material
bulk.\cite{grain_bound1968,uphill1968-2,uphill0,cowern2007} It
has been shown that segregation of the impurities near the surface
produced by the diffusion of bound i-d pairs has some peculiarities
that may have important practical consequences. The spatial scale
at which the peculiarities are the most pronounced is determined by
the average migration distance of the bound i-d pair $\lambda$, which
experimentally was found to be in the nanometer range in semiconductor
materials\cite{cowern1990,cowern1992,cowern2007} and theoretically
estimated to have similar values in metals.\cite{I}  At the macroscopic
scale much exceeding $\lambda$ it does not matter whether the elementary
diffusion step is of the order of the lattice constant or is equal
to $\lambda$.  But if the length scale of interest is similar to
$\lambda$, both quantitative and qualitative differences can be detected.
The underlying cause of the differences is that at the microscopic scale
the pair diffusion does not obey Fick's laws\cite{cowern1990} which
is reflected in the nonlocal nature of the diffusion current.  The n-F
features originate from the permanent presence of the diffusion mediating
impurity within the pair which makes possible its autonomous migration at
$O(\lambda)$ distances independently of the ambient defect and impurity
concentrations.\cite{cowern1990,mirabella2009} This, however, takes place
only during the pair lifetime which restricts the n-F behavior to the
length scale of $O(\lambda)$. The latter is temperature dependent and at
low temperatures may grow on orders of magnitude\cite{cowern1992} which
can be used as an additional means of control of the segregation profiles.

Experimentally the difference between the profiles produced by the pair
diffusion and by the mechanisms that admit description by local diffusion
equations\cite{irradiation1976,Wiedersich83,SOISSON2006235,drag-inBCC}
can be observed only at early stages of evolution because at late stages
both mechanisms lead to identical stationary and/or equilibrium profiles.
It should be stressed that the stationary defect distributions have been
used in the present paper only for simplicity and that the underlying
microscopic mechanism should be operative for time-dependent distributions
as well. Short-lived supersaturation can be even advantageous because
it may be used to arrest segregation at an early stage to obtain a
concentration profile of desired width.

The most promising experimental setup for verification of the predictions
about the pair drag in semiconductors seems to be similar to that used
in Ref.\ \onlinecite{cowern2007} in the study of the uphill diffusion
only the initial impurity distribution near the surface should be much
flatter in order to facilitate the measurement of the small (due to
the short diffusion time) differences between the profile evolution
predicted in different theories, such as those discussed in Ref.\
\onlinecite{cowern2007}.

The stationary irradiation conditions in metallic systems
\cite{irradiation1976,Wiedersich83,SOISSON2006235,radiation2016} are
not well suited to investigation of the nonlocality effects, in particular,
because of the presence of several modes of diffusion propagated by different
defects. But the quench experiments similar to those of Refs.\
\onlinecite{grain_bound1968,uphill1968-2} for solutes with strong
binding to vacancies\cite{I} could be suitable for the purpose except
that the spatial extent of the segregation profiles should be reduced
to the nanoscale range.\cite{radiation2016} A transient nature of the
supersaturation in such experiments may turn out to be even advantageous
because the pair contribution is best visible at the early stages when
only a few i-d migration events occurred and this number can be controlled
by the strength and the duration of the non-equilibrium defect flux.

In conclusion it should be noted that the phenomenological theory of
the pair diffusion developed in the present paper is only a particular
case of more general notion of the mobile state suggested in Ref.\
\onlinecite{cowern1990}.  The mobile state of impurity should not
necessary be associated with the pairing but may be caused, for example,
by the kick-out mechanism.\cite{cowern1990,cowern1992} Other possibilities
can be envisaged and if the transition to the mobile state is driven by
the defects the formalism developed in the present paper should apply
because the pairs have been assumed to be unstructured unstable point
particles that can be associated, for example, with the interstitial state
invoked in Refs.\ \onlinecite{irradiation1976,Wiedersich83,SOISSON2006235}
to describe the radiation-induced segregation. The interstitials are
unstable due to recombination processes so if the value of the mean
migration path $\lambda$ is of appreciable value their isolated motion at
distances of order $\lambda$ can be described by the nonlocal diffusion
equations derived in the present paper.  In this respect the results
of the present paper may be viewed as an alternative approach to the
description of the diffusion via a fast intermediate suggested in Refs.\
\onlinecite{cowern1990,general_expression2010}.
\begin{acknowledgments}
I am grateful to Ren\'e Monnier for carefully reading a preliminary
version of the manuscript and for useful suggestions on its
improvement. Also, I would like to express my gratitude to Hugues
Dreyss\'e for support and encouragement.
\end{acknowledgments}
\appendix
\section{\label{appA}Comparison with the diffusion profile of Ref.\ 
\onlinecite{cowern1990}}
In slightly modified notation, the evolution of 1D impurity profile
starting from the delta-function distribution at $t=0$ can be described
by Eqs.\ (5)-(10) of Ref.\ \onlinecite{cowern1990} as
\begin{equation}
C(\xi,\theta)=\lambda^{-1}\sum_{n=0}^\infty P_n(\theta)\phi_n(\xi,1),
	\label{C(xi)}
\end{equation}
where
\begin{equation}
	\theta = gt,\qquad \xi = x/\lambda,
	\label{xi-theta}
\end{equation}
\begin{equation}
	P_n(\theta)=(\theta^n/n!)e^{-\theta},
	\label{Pn}
\end{equation}
\begin{equation}
	\phi_{n=0}(\xi,1)=\delta(\xi),
	\label{fn=0}
\end{equation}
and
\begin{equation}
\phi_{n>0}(\xi,\nu)=\frac{e^{-\sqrt{\nu}|\xi|}}{(2\sqrt{\nu})^{2n-1}}
\sum_{k=0}^{n-1}\frac{2^k}{k!}\binom{2n-2-k}{n-1}(|\xi|\sqrt{\nu})^k.
\label{fn>0}
\end{equation}
Here we introduced the parameter $\nu$ to facilitate the proof that
the profile Eq.\ (\ref{C(xi)}) of Ref.\ \onlinecite{cowern1990}
coincides with the 1D profile obtained from Eq.\ (\ref{Ckt}) by the inverse
Fourier transform with $C_0(k)=1$ corresponding to
the delta-function initial profile:
\begin{eqnarray}
C(x,t)&=&\frac{1}{2\pi}\int_{-\infty}^\infty dk e^{ixk} e^{-gt}\exp\left(\frac{gt}{1+(\lambda k)^2}\right)\nonumber\\
&=&\frac{1}{\lambda}
\sum_{n=0}^\infty P_n(\theta) \frac{1}{2\pi}\int_{-\infty}^\infty 
\frac{e^{i\xi u}du}{(1+ u^2)^n}.
	\label{C(X)}
\end{eqnarray}
Here the last exponential on the first line has been expanded in the
Tailor series so by comparison with Eq.\ (\ref{C(xi)}) we conclude that
the inverse Fourier transforms on the second line should be equal to
$\phi_{n>0}(\xi,1)$. To show this we first introduce the integrals
\begin{equation}
	\phi_{n}(\xi,\nu)=\frac{1}{2\pi}\int_{-\infty}^\infty 
	\frac{e^{i\xi u}du}{(\nu+ u^2)^n}
	\label{phi(mu)}
\end{equation}
and note that if $\phi_1(\xi,\nu)$ is known explicitly, other integrals
can be computed recursively as
\begin{equation}
	\phi_{n+1}(\xi,\nu)=-\frac{1}{n}\frac{d}{d\nu}\phi_{n}(\xi,\nu).
	\label{n+1}
\end{equation}
Thus, we only need to show that $\phi_{n}(\xi,\nu)$ in Eq.\ 
(\ref{fn>0}) satisfy the recursion.  To this end we first note that
with the exponential factor being common to all terms in all functions,
the equality in Eq.\ (\ref{n+1}) will hold if it will be valid for every
power of $|\xi|^k$ under the summation sign. Let us consider one such
term in Eq.\ (\ref{fn>0})
\begin{equation}
	\phi_n^{(k)}=\frac{e^{-\sqrt{\nu}|\xi|}}{2^{2n-1}(n-1)!}
	\frac{2^k}{k!}\frac{(2n-2-k)!}{(n-1-k)!}|\xi|^k\nu^{(k+1)/2-n}.
	\label{phink}
\end{equation}
When substituted in Eq.\ (\ref{n+1}) it will contribute to $|\xi|^k$
term in $\phi_{n+1}^{(k)}$ through the derivative of its last factor with
respect to $\nu$. The only other contribution from $\phi_n$ contributing
into $|\xi|^k$ term in $\phi_{n+1}$ is $\phi_n^{(k-1)}$ differentiated
with respect to $\nu$ in the exponential function. It is straightforward
to check that these two contributions lead to the term $\phi_{n+1}^{(k)}$
as in Eq.\ (\ref{phink}) only with $n+1$ instead of $n$, as required.
\section{\label{app}Diffusion near the boundary in 1D geometry}
If only segregation near the surface is of interest and $\lambda\ll L$
(in the examples in the main text $\lambda=5-10$~nm and $L=200$~nm), at
small $x$ the influence of the second surface of the film
at $x=2L$ can be neglected and the system can be considered as infinite
in the direction of $x>0$.  In this case the diffusion kernel in
the half-space $x,x_0\geq 0$ can be found by the method of images as
\begin{equation}
	P(x,x_0)=\frac{1}{2\lambda}\left[e^{-|x-x_0|/\lambda}
		+e^{-(x+x_0)/\lambda}\right].
	\label{Pxx0}
\end{equation}
It is straightforward to check that (\ref{Pxx0}) satisfies both equation
(\ref{P_eq}) and the boundary condition (\ref{BC}) at $x=0$.

Even in 1D systems with the simple stationary pair creation rate
(\ref{g(x)}) the diffusion equation (\ref{the_eq}) cannot be solved
analytically for all $t$. However, at the beginning of the evolution when
the impurity distribution still has its simple constant initial form Eq.\
(\ref{C-const}) the solution can be found to the first order in $t$
as the sum of the initial distribution and the integral on the r.h.s.\
of Eq.\ (\ref{the_eq}) multiplied by $t$.\cite{cowern1990}  Substituting
the constant $C_{i0}$, Eqs.\ (\ref{g(x)}) and (\ref{Pxx0}) in Eq.\
(\ref{the_eq}) one gets after simple integrations 
\begin{equation}
C_i(x,t)\approx C_{i0}\left(1+tg_s\frac{\lambda}{d_s}e^{-x/\lambda}\right)
\label{Cxt=0}
\end{equation}
up to a small correction of $O(1/s)$.
\providecommand{\newblock}{}
\end{document}